\documentclass{elsart}
\usepackage{amssymb}
\usepackage{amsmath}
\usepackage{graphicx}

\journal{Physics Letters A}

\begin{document}

\begin{frontmatter}

\title{Lower limit in semiclassical form for the number of bound states in a central potential}

\author{Fabian Brau\thanksref{fnrs}}
\thanks[fnrs]{FNRS Postdoctoral Researcher}
\address{Service de Physique G\'en\'erale et de Physique des Particules El\'ementaires, Groupe de Physique Nucl\'eaire Th\'eorique, Universit\'e de Mons-Hainaut, Mons, Belgique}
\ead{fabian.brau@umh.ac.be}

\author{Francesco Calogero}
\address{Dipartimento di Fisica, Universit\`a di Roma ``La Sapienza" and Istituto Nazionale di Fisica Nucleare, Sezione di Roma, Rome, Italy}
\ead{francesco.calogero@roma1.infn.it}

\date{\today}

\begin{abstract}
We identify a class of potentials for which the semiclassical estimate $N^{\text{(semi)}}=\frac{1}{\pi}\int_0^\infty dr\sqrt{-V(r)\theta\left[-V(r)\right]}$ of the number $N$ of (S-wave) bound states provides a (rigorous) lower limit: $N\ge \left\{\left\{N^{\text{(semi)}}\right\}\right\}$, where the double braces denote the integer part. Higher partial waves can be included via the standard replacement of the potential $V(r)$ with the effective $\ell$-wave potential $V_\ell^{\text{(eff)}}(r)=V(r)+\frac{\ell(\ell+1)}{r^2}$. An analogous upper limit is also provided for a different class of potentials, which is however quite severely restricted.
\end{abstract}

\end{frontmatter}


\section{Introduction and main results}

\label{sec1}

The number $N$ of (S-wave) bound states possessed, in the framework of nonrelativistic quantum mechanics in ordinary (three-dimensional) space, by a central potential $V(r)$, coincides with the number of zeros, in the interval $0<r<\infty $, of the solution of the zero-energy (S-wave) radial Schr\"{o}dinger equation 
\begin{subequations}
\label{eq1}
\begin{equation}
\label{eq1a}
u''(r)=V(r)\,u(r)  
\end{equation}
characterized by the boundary condition 
\begin{equation}
\label{eq1b}
u(0)=0.  
\end{equation}
\end{subequations}
Here and throughout appended primes denote differentiations and we use units such that $\hbar =2m=1$, where $m$ is the mass of the particle bound by the potential $V(r)$, which is hereafter assumed to have the standard properties generally required in this context: to vanish at infinity, and to be such that the integral written in the following formula (\ref{eq2}) is finite. The ``semiclassical" estimate $N^{\text{(semi)}}$ for $N$ reads (see, for instance, \cite{tom}; but note that $N^{\text{(semi)}}$ is generally not an integer) 
\begin{equation}
\label{eq2}
N^{\text{(semi)}}=\frac{1}{\pi}\int_0^{\infty}dr\,\sqrt{-V^{(-)}(r)}.
\end{equation}
Here and throughout we use the notation $V^{(-)}(r)$ to denote the negative (``attractive") part of the corresponding potential $V(r)$, 
\begin{equation}
\label{eq3}
V^{(-)}(r)=V(r)\,\theta \left[-V(r)\right].  
\end{equation}
Here and below $\theta (x)$ denotes the standard step function, $\theta(x)=0$ if $x<0$, $\theta (x)=1$ if $x\geq 0$. These results, as well as those discussed in the rest of this paper, can be extended to higher partial waves characterized by the angular momentum quantum number $\ell$, via the standard replacement of the potential $V(r)$ with the ``effective" $\ell$-wave potential 
\begin{equation}
\label{eq4}
V_{\ell }^{\text{(eff)}}(r)=V(r)+\frac{\ell (\ell +1)}{r^2}.
\end{equation}
In 1968 Chadan \cite{cha} has shown that -- consistently with the ``correspondence principle" relating quantum mechanics at
large quantum numbers with classical mechanics -- the number $N$ of S-wave bound states (as well as the number of bound states for any fixed angular momentum $\ell$) possessed by the potential 
\begin{equation}
\label{eq5}
V(r)=g^2v(r)  
\end{equation}
grows asymptotically, when the strength $g^2$ of the potential diverges, just as the semiclassical estimate (\ref{eq2}): 
\begin{equation}
\label{eq6}
N\approx \frac{g}{\pi }\int_{0}^{\infty }dr\,\sqrt{-v^{(-)}(r)}=\frac{1}{\pi}\int_{0}^{\infty}dr\,\sqrt{-V^{(-)}(r)}=N^{{\text{(semi)}}}\quad \text{as}\quad g\rightarrow \infty .  
\end{equation}
Here the symbol $\approx$ denotes asymptotic equality (up to lower order additive corrections). This finding entails of course that, for strong potentials possessing many bound states, the semiclassical estimate (\ref{eq2}) provides a good approximation to the number of (S-wave) bound states $N$. More recently (rigorous) upper and lower limits on the number $N$ of (S-wave) bound states have been obtained \cite{bra1,bra2}, which are mainly given by the semiclassical estimate $N^{\text{(semi)}}$ (\ref{eq2}), and that therefore generally approximate well the (large) number of bound states
possessed by strongly binding potentials; but these limits also feature certain additional terms, that generally only play a minor quantitative role (when the number of bound states is not too small), yet mar the neatness of these results inasmuch as they introduce a not-too-transparent dependence on the potential $V(r)$. These limits have been proven in \cite{bra1} for
monotonically increasing potentials, and have been generalized in \cite{bra2} to more general potentials, also in order to include the higher partial wave case (note that the effective $\ell$-wave potential (\ref{eq4}) cannot belong to the class of monotonically increasing potentials). They read, for monotonic potentials, 
\begin{subequations}
\label{eq7}
\begin{equation}
\label{eq7a}
N<N^{\text{(semi)}}+\frac{1}{4\pi}\log \left\vert \frac{V(p)}{V(q)}\right\vert +\frac{1}{2},  
\end{equation}
\begin{equation}
\label{eq7b}
N>N^{\text{(semi)}}-\frac{1}{4\pi }\log \left\vert \frac{V(p)}{V(q)}\right\vert -\frac{3}{2},  
\end{equation}
where we use of course the definition (\ref{eq2}) of $N^{\text{(semi)}}$ and the two radii $p$ and $q$ are defined by the two relations 
\begin{equation}
\label{eq7c}
\int_0^p dr\,\sqrt{-V^{(-)}(r)}=\frac{\pi }{2}\quad \text{and}\quad \int_q^{\infty}dr\,\sqrt{-V^{(-)}(r)}=\frac{\pi}{2}.  
\end{equation}
Purpose and scope of this paper is to identify a class of potentials for which the semiclassical expression (\ref{eq2}) provides itself a (rigorous) lower limit for the number $N$ of S-wave bound states -- with the possibility to extend these results to higher partial waves via the replacement of the potential $V(r)$ with the effective $\ell$-wave potential (\ref{eq4}). We also identify below (see Remark 2) another class of potentials for which the semiclassical estimate (\ref{eq2}) provides an upper bound to the number of S-wave bound states -- but this class is so much more restricted (including the fact that it does not allow the extension to higher partial waves) that we decided to focus this paper on
the lower limit (see title).

To obtain our neat lower bound we restrict attention to potentials that possess at most two zeros in the interval $0<r<\infty$ and that are negative (``attractive") between them: 
\end{subequations}
\begin{subequations}
\label{eq8}
\begin{equation}  \label{eq8a}
V(r_-)=V(r_+)=0,
\end{equation}
\begin{equation}  \label{eq8b}
V(r) < 0 \quad \text{for} \quad r_-<r<r_+,
\end{equation}
\begin{equation}  \label{eq8c}
V(r) > 0 \quad \text{for} \quad r<r_- \quad \text{and for} \quad r_+<r.
\end{equation}
This restriction is generally adequate to accommodate most cases of interest -- including the treatment of higher partial waves via the replacement of the potential $V(r)$ with the effective $\ell$-wave potential (\ref{eq2}). Note that we are not excluding the possibility that $r_-$ not be positive (namely, potentials attractive rather than repulsive at the origin) or $r_+$ be infinity (namely, potentials attractive rather than repulsive at infinity) -- in which cases the potential would possess less than two zeros in the interval $0<r<\infty$, and (\ref{eq8}) should be modified accordingly. As for the upper bound result, it only applies to everywhere attractive potentials without any zeros, and with additional restrictions,
see Remark 2 below.

We now state our main result. Consider, in the interval $r_-<r<r_+$, the auxiliary function $F(r)$ defined, in terms of the original potential $V(r)$, as follows: 
\end{subequations}
\begin{equation}
\label{eq9}
F(r)=\frac{5}{16}\left[ \frac{V'(r)}{V(r)}\right] ^{2}-\frac{V''(r)}{4V(r)}.  
\end{equation}
Note that this auxiliary function does not depend on the strength $g^2$ of the potential, see (\ref{eq5}), but only on its shape, and that it is finite inside the interval $r_-<r<r_+$, although it generally diverges to positive infinity at its borders, see (\ref{eq8a}). Our class of potentials is then characterized by the property that this function $F(r)$ be positive not only at the borders of the interval $r_-<r<r_+$, but as well that it be nonnegative throughout this interval: 
\begin{equation}
\label{eq10}
F(r)\geq 0\quad \text{for}\quad r_-<r<r_+.
\end{equation}
For this class of potentials there holds then the following (rigorous) lower limit on the number $N$ of S-wave bound states: 
\begin{equation}
\label{eq11}
N\geq \left\{ \left\{ N^{\text{(semi)}}\right\} \right\} .  
\end{equation}
Here and throughout the double braces denote the integer part. This results is proven in Section \ref{sec3}, so as to make this paper self-contained (actually this finding is an extension of a result obtained in \cite{bra2}, see the Remark 1 below). And the remarkable stringency of the lower limit (\ref{eq11}) is demonstrated in the following Section \ref{sec2}, for various test potentials. 

We end this section with 6 remarks.

\textit{Remark 1}. In \cite{bra2} a treatment analogous to that reported in Section \ref{sec3} below was given in the more general context of $\ell$-waves (but the potential was supposed to be negative for all values of $r$, we show here that this restriction is not necessary); here we restrict our consideration to the S-wave case, because our main purpose in this paper is to exhibit the neat lower limit (\ref{eq11}) closely related to the semiclassical expression (\ref{eq2}). The possibility remains of course to apply this result to higher partial waves via the replacement of the potential $V(r)$ with the effective $\ell $-wave potential (\ref{eq4}).

\textit{Remark 2}. As implied by the treatment given in Section \ref{sec3}, for the class of potentials characterized by the condition opposite to (\ref{eq10}), 
\begin{equation}  
\label{eq12}
F(r) \le 0,
\end{equation}
there holds the upper limit 
\begin{equation}  
\label{eq13}
N \le N^{\text{(semi)}}-1.
\end{equation}
But, due to the fact that the auxiliary function $F(r)$ generally diverges to positive infinity where the potential $V(r)$ vanishes, this class can only include potentials without any zero (namely, everywhere attractive -- thereby excluding the extension of the result to higher partial waves via the effective potential (\ref{eq4})), and it requires moreover that the
potential $V(r)$ vanish asymptotically, as $r \rightarrow \infty$, no faster than $r^{-4}$ (to prevent the auxiliary function $F(r)$ from becoming positive as $r \rightarrow \infty$, thereby violating (\ref{eq12}) which should of course now hold for all positive values of $r$).

\textit{Remark 3}. For the square-well potential 
\begin{subequations}
\label{eq14}
\begin{equation}
\label{eq14a}
V(r)=-g^2 R^{-2}\,\theta (R-r)  
\end{equation}%
(for which the function $F(r)$ vanishes trivially in the interval $0\leq
r<r_+=R$), 
\begin{equation}
\label{eq14b}
N=\left\{ \left\{ \frac{g}{\pi }+\frac{1}{2}\right\} \right\},
\end{equation}
and 
\begin{equation}
\label{eq14c}
N^{\text{(semi)}}=\frac{g}{\pi }.  
\end{equation}
Hence in this case the lower limit (\ref{eq11}) is essentially saturated (as well as the upper limit (\ref{eq13})). The slight discrepancy is due to the fact that, for $r>R$, (\ref{eq14a}) does not quite satisfy (8c).

\textit{Remark 4}. The special case of a potential $V(r)$ such that the auxiliary function (\ref{eq9}) vanishes (nontrivially), $F(r)=0$, has been discussed separately (in fact, in a more general context) \cite{bra3}.

\textit{Remark 5}. Let us emphasize the crucial role played by the ``shape" conditions (\ref{eq8}) and (\ref{eq10}) (or (\ref{eq12})). It is indeed easy to show that there exist potentials -- obviously not restricted by these conditions -- that possess no bound states at all, $N=0$, while the corresponding value of $N^{\text{(semi)}}$ is arbitrarily large, as well
indeed as potentials that possess an arbitrarily large number of bound states $N$ while the corresponding value of $N^{\text{(semi)}}$ is arbitrarily small; for instance a potential of this second kind can be realized as an appropriate sequence of negative delta functions, while a potential of the first kind -- contradicting dramatically the bound (\ref{eq11}) -- can be realized as an arbitrarily long negative square well with an appropriate sequence of positive delta functions embedded in it.

\textit{Remark 6}. Finally, let us note that, due to the ease nowadays to compute numerically the number of bound states for any given potential (especially using techniques such as those described in \cite{cal}), the results reported in this paper have mainly an academic -- rather than a practical -- relevance; except in the case of strongly binding potentials possessing very many bound states, where numerical computations might be somewhat cumbersome, while the rigorous bounds reported above might yield explicitly computable results that are moreover likely to be quite close to the exact results (see some of the examples in the following Section \ref{sec2}).

\section{Tests}

\label{sec2}

We test in this section the new lower limit (\ref{eq11}) with some specific potentials. The first potential we consider is the (solvable) Morse potential \cite{mor} 
\end{subequations}
\begin{equation}
\label{eqt1}
V(r)=-g^{2}R^{-2}\left\{ 2\exp \left[ -\frac{r}{R}+\alpha \right] -\exp\left[ -2\frac{r}{R}+2\alpha \right] \right\} ,  
\end{equation}
where $\alpha$, as well of course as $R$, is an arbitrary positive constant. This potential has a single zero at $r=R\,(\alpha -\log 2)$ if $\alpha \geq \log 2,$ otherwise it is negative (attractive) in the entire interval $0\leq r<\infty$. The number of its (S-wave) bound states $N$ turns out to be independent of $\alpha $:
\begin{equation}
\label{eqt2}
N=\left\{ \left\{ g+\frac{1}{2}\right\} \right\} .  
\end{equation}
For this potential, $F(r)$ is positive (for all values of $\alpha$) hence the lower limit (\ref{eq11}) applies and it reads
\begin{equation}
\label{eqt3}
N\geq \left\{ \left\{ g\right\} \right\} .  
\end{equation}
Thus the maximal gap between the exact result and the lower bound is, at most, of one unit for all values of $g$.

The second potential we consider is the (solvable, and everywhere negative) P\"{o}schl-Teller \cite{PT} -- or ``single
soliton" (see for instance \cite{CD}) -- potential 
\begin{equation}
\label{eqt4}
V(r)=-g^2 R^{-2}\,\left[ \cosh \left( \frac{r}{R}\right) \right]^{-2}.
\end{equation}
The number $N$ of (S-wave) bound states for this potential is 
\begin{equation}
\label{eqt5}
N=\left\{ \left\{ \frac{1}{4}\left( 1+\sqrt{1+4g^2}\right) \right\}
\right\} .  
\end{equation}
For this potential, $F(r)$ is also everywhere positive hence the lower limit (\ref{eq11}) applies and it reads
\begin{equation}
\label{eqt6}
N\geq \left\{ \left\{ \frac{g}{2}\right\} \right\} .  
\end{equation}
Thus the maximal gap between the exact result and the lower bound is again, at most, of one unit for all values of $g$.

The third potential we consider is the Lennard-Jones potential 
\begin{equation}
\label{eqt7}
V(r)=g^{2}R^{-2}\,\left[ \left( \frac{R}{r}\right)^{12}-\left( \frac{R}{r}\right) ^{6}\right],  
\end{equation}
which clearly has a single zero at $r=R$. In this case, the exact number $N$ of (S-wave) bound states is not computable analytically, hence numerical calculations are necessary. For this potential, $F(r)$ is also positive. The lower limit (\ref{eq11}), which can be computed analytically, reads
\begin{equation}
\label{eqt8}
N\geq \left\{ \left\{ \frac{g}{12\sqrt{\pi}}\,\frac{\Gamma(1/3)}{\Gamma(11/6)}\right\} \right\}\cong \left\{ \left\{ 0.1339\,g\right\} \right\} .
\end{equation}
Numerical investigations for $0<g\leq 500$ ($g=500$ yields 67 bound states) show that the maximal gap between the exact result and the lower bound is again, at most, of one unit for all these values of $g$.

The last (everywhere attractive class of) potential(s) we consider reads
\begin{equation}
\label{eqt9}
V(r)=-g^2 R^{-2}\,\left( \frac{r}{R}\right) ^{\alpha -2}\,\exp \left[-\left( \frac{r}{R}\right) ^{\beta }\right]   
\end{equation}
where $\alpha $ and $\beta $ are two arbitrary positive constants, $\alpha >$ $\beta >0$, that satisfy the following condition: 
\begin{equation}
\label{eqt10}
\alpha \beta \geq \beta ^{2}+1.  
\end{equation}
This inequality is necessary and sufficient to guarantee validity of the inequality (\ref{eq10}) (with $r_-=0$, $r_+=\infty $), hence the applicability of the lower limit (\ref{eq11}), which can be evaluated exactly and it yields the explicit lower limit 
\begin{equation}
\label{eqt11}
N\geq \left\{ \left\{ \frac{g}{\pi \beta }\,2^{\frac{\alpha }{2\beta }}\,\Gamma \left( \frac{\alpha }{2\beta }\right) \right\} \right\} .
\end{equation}
In particular, when $\alpha =2$ and $\beta =1$, we obtain the lower limit $N\geq \{\{2g/\pi \}\}$ on the number of S-wave bound states for the exponential potential $V(r)=-g^2 R^{-2}\,\exp \left( -\frac{r}{R}\right) $, which simplifies and improves the lower limit given in our previous work (see eq. (2.13) of Ref. \cite{bra1}). Numerical investigations for this
exponential potential with $0<g\leq 200$ ($g=200$ yields 127 bound states) show that the maximal gap between the exact result and the lower bound is again, at most, of one unit for all these values of $g$.

\section{Proof}

\label{sec3}

In this section we prove our main result, as reported in Section \ref{sec1}. Our main task is to count -- or rather bound from below -- the number $N$ of zeros (in the interval $0<r<\infty $) of the wave function $u(r)$ characterized by (\ref{eq1}), with a potential $V(r)$ that has, to begin with, the property (\ref{eq8}). The function $u(r)$ has no zeros for $0<r\leq r_-$, since it vanishes at the origin, see (\ref{eq1b}), and it is convex in the interval $0<r<r_-$ (see (\ref{eq1a}) and (\ref{eq8c})); and it can possess at most one zero in the region $r_{+}<r<\infty $ where it is also convex (see (\ref{eq1a}) and (\ref{eq8c})). Therefore to bound from below the number $N$ of its zeros it is sufficient to consider the inner interval $r_{-}<r<r_{+}$. To count the zeros of $u(r)$ in this interval it is convenient to introduce the function $\eta (r)$ by setting 
\begin{equation}
\label{eq15}
\sqrt{-V(r)}\cot \left[ \eta (r)\right] =\frac{u'(r)}{u(r)}+\frac{V'(r)}{4V(r)}.  
\end{equation}
It is then easily seen, via (\ref{eq1}), that $\eta (r)$ satisfies the first-order equation 
\begin{equation}
\label{eq16}
\eta'(r)=\sqrt{-V(r)}+\frac{F(r)}{\sqrt{-V(r)}}\sin ^{2}\left[\eta (r)\right]   
\end{equation}
where we used the definition (\ref{eq9}), and this implies 
\begin{equation}
\label{eq17}
\eta(r_+)-\eta(r_-)=\int_{r_-}^{r_+}dr\,\sqrt{-V(r)}+\int_{r_-}^{r_+}dr\,\frac{F(r)}{\sqrt{-V(r)}}\sin^2\left[\eta(r)\right], 
\end{equation}
hence, via (\ref{eq10}), 
\begin{subequations}
\label{eq18}
\begin{equation}
\label{eq18a}
\eta(r_+)-\eta(r_-)>\int_{r_-}^{r_+}dr\,\sqrt{-V(r)},
\end{equation}
hence, via (\ref{eq8c}) and (\ref{eq3}), 
\begin{equation}
\label{eq18b}
\eta(r_+)-\eta(r_-)>\int_{0}^{\infty }dr\,\sqrt{-V^{(-)}(r)},
\end{equation}
hence, via (\ref{eq2}), 
\begin{equation}
\label{eq18c}
\eta(r_+)-\eta(r_-)>\pi N^{\text{(semi)}}.  
\end{equation}
In these last three formulas, (\ref{eq18}), we used the strict inequality sign, neglecting for simplicity the very marginal cases when this would not be justified. We now observe that the differential equation (\ref{eq16}) implies that, every time $\eta (r)$ goes through an integer multiple of $\pi$, its derivative $\eta'(r)$ is positive, while the formula (\ref{eq15}) implies that, every time $\eta(r)$ goes through an integer multiple of $\pi$, the wave function $u(r)$ vanishes. It is moreover clear from (\ref{eq15}) and (\ref{eq8a}) that both $\eta(r_-)$ and $\eta(r_+)$ are integer multiples of $\pi$; a result which can be obtained as well from the differential equation (\ref{eq16}) (integrate it forward from a value just
before $r_+$ to $r_+,$ or backward from a value just after $r_-$ to $r_-,$ taking into account the divergence of $F(r)$ at $r_+$ and at $r_-,$ see (\ref{eq9}) and (\ref{eq8a})) . One therefore concludes (sketch a graph of $\eta(r)$ using the above information!) that the number $\tilde{N}$ of zeros of $u(r)$ in the interval $r_-<r<r_+$ is given by the expression 
\end{subequations}
\begin{equation}
\label{eq19}
\tilde{N}=\frac{\eta (r_{+})-\eta (r_{-})}{\pi }-1.  
\end{equation}
(Here we exclude from consideration the marginal case in which $u(r_+)=0)$. We now note that, if $r_+=\infty$, all the zeros of the wave function $u(r)$ for $0<r<\infty $ are in this interval, hence in this case their number $N$ coincides with $\tilde{N}.$ If instead $r_+ <\infty $, the last zero of the wave function can be inside respectively outside the interval $r_- <r<r_+$, yielding $N=\tilde{N}$ respectively $N=\tilde{N}+1$. In any case we conclude that $N\geq \tilde{N}$, and this, together with (\ref{eq19}) and (\ref{eq18c}), entails (\ref{eq11}). Q. E. D.

The modification of this proof to validate the upper limit (\ref{eq13}) is obvious: all one needs to note is that the replacement of (\ref{eq10}) with (\ref{eq12}) entails that the inequalities (\ref{eq18}) must be reversed, and then via the analysis just made above (the case $r_+<\infty$ necessarily violate (\ref{eq12})) one gets (\ref{eq13}). Q. E. D.

\end{document}